\documentclass[9pt, sigconf]{acmart}

\usepackage{balance}
\usepackage{enumitem,kantlipsum}
\usepackage{soul}
\usepackage[linesnumbered,ruled]{algorithm2e}
\usepackage{amsmath,amsfonts}
\usepackage{algorithmic}
\usepackage{graphicx}
\usepackage{textcomp}
\usepackage{xcolor}
\usepackage{subcaption}
\usepackage{titlesec}

\definecolor{mintbg}{rgb}{.63,.79,.95}

\newcommand{\create}{\mathsf{CREATE}}
\newcommand{\transfer}{\mathsf{TRANSFER}}

\newcommand{\scdb}{\textnormal{S\kern-.1em\textsc{mart}\textsc{chain}DB}}
\newcommand{\request}{\mathsf{REQUEST}}
\newcommand{\bidcap}{\mathsf{BID}}
\newcommand{\acceptbid}{\mathsf{ACCEPT\_BID}}
\newcommand{\returnbid}{\mathsf{RETURN}}

\newcommand{\server}{\mathit{Server}}
\newcommand{\driver}{\mathit{Driver}}
\newcommand{\mongodb}{\mathit{MongoDB}}
\newcommand{\tendermint}{\mathit{Tendermint}}

\colorlet{lightmintbg}{mintbg!40}

\newcommand{\bigchaindb}{\textnormal{B\kern-.1em\textsc{ig}\textsc{chain}DB}}
\newcommand{\smartchaindb}{\textnormal{S\kern-.1em\textsc{mart}\textsc{chain}DB}}

\newcommand{\ethereum}
{\textnormal{E\kern-.1em\textsc{thereum}}}
\newcommand{\hyper}{\textnormal{H\textsc{yperledger}}\textnormal{ F\textsc{abric}}}
\newcommand{\bitcoinTuenned}{\textnormal{B\kern-.1em\textsc{itcoin}}}

\newcommand{\definition}{\textbf{D\textsc{efinition}}}
\newcommand{\example}{\textnormal{E\textsc{xample}}}

\setcopyright{none}
\renewcommand\footnotetextcopyrightpermission[1]{}

\setlist[itemize]{leftmargin=*}
\setlist[enumerate]{leftmargin=*}
\titlespacing*{\section}{2pt}{4pt}{2pt}
\titlespacing*{\subsection}{2pt}{4pt}{2pt}
\titlespacing*{\subsubsection}{2pt}{6pt}{2pt}

\AtBeginDocument{%
  \providecommand\BibTeX{{%
    \normalfont B\kern-0.5em{\scshape i\kern-0.25em b}\kern-0.8em\TeX}}}

\setcopyright{acmlicensed}
\copyrightyear{2018}
\acmYear{2018}
\acmDOI{XXXXXXX.XXXXXXX}

\acmConference[]{}{}{}
\acmISBN{978-1-4503-XXXX-X/18/06}

\begin{document}

\title{Taming the Beast of User-Programmed Transactions on Blockchains: A Declarative Transaction Approach}

\author{Nodirbek Korchiev}
\affiliation{%
  \institution{North Carolina State University}
  \streetaddress{890 Oval Dr}
  \city{Raleigh}
  \state{North Carolina}
  \country{USA}
  \postcode{27695}
}
\email{nkorchi@ncsu.edu}

\author{Akash Pateria}
\affiliation{%
  \institution{Oracle}
  \city{Seattle}
  \state{Washington}
  \country{USA}
  \postcode{27695}
}

\email{pateria.akash77@gmail.com}
\author{Vodelina Samatova}
\affiliation{%
  \institution{North Carolina State University}
  \streetaddress{890 Oval Dr}
  \city{Raleigh}
  \state{North Carolina}
  \country{USA}
  \postcode{27695}
}
\email{vsamato@ncsu.edu}

\author{Sogolsadat Mansouri}
\affiliation{%
  \institution{North Carolina State University}
  \streetaddress{890 Oval Dr}
  \city{Raleigh}
  \state{North Carolina}
  \country{USA}
  \postcode{27695}
}
\email{smansou2@ncsu.edu}

\author{Kemafor Anyanwu}
\affiliation{%
  \institution{North Carolina State University}
  \streetaddress{890 Oval Dr}
  \city{Raleigh}
  \state{North Carolina}
  \country{USA}
  \postcode{27695}
}
\email{kogan@ncsu.edu}

\renewcommand{\shortauthors}{Korchiev et al.}

\begin{abstract}
 Blockchains are being positioned as the "technology of trust" that can be used to mediate \textit{transactions} between non-trusting parties without the need for a central authority. They support transaction types that are \textit{native} to the blockchain platform or \textit{user-defined} via user programs called \textit{ smart contracts}. Despite the significant flexibility in transaction programmability that smart contracts offer, they pose several usability, robustness and performance challenges. 

This paper proposes an alternative transaction framework that incorporates more primitives into the native set of transaction types (reducing the likelihood of requiring user-defined transaction programs often). The framework is based on the concept of \textit{declarative blockchain transactions} whose strength lies in the fact that it addresses several of the limitations of smart contracts, simultaneously. A formal and implementation framework is presented and a subset of commonly occurring transaction behaviors are modeled and implemented as use cases, using an open-source blockchain database, $\bigchaindb$ as the implementation context. A performance study comparing the declarative transaction approach to equivalent smart contract transaction models reveals several advantages of the proposed approach.

\end{abstract}



\received{20 February 2007}
\received[revised]{12 March 2009}
\received[accepted]{5 June 2009}

\maketitle

\section{Introduction}
Blockchains, as a technology for mediating and managing transactions between non-trusting parties, is becoming an increasingly popular concept. They are decentralized, fully replicated, append-only databases of \textit{transactions} that are \textit{validated} through a large, distributed consensus. These characteristics ensure that blockchain contents are tamper-proof and that no single authority controls a blockchain's operation and contents, conferring a good degree of trust in them. 

 


Initially aimed at cryptocurrency, blockchain technology now extends to areas seeking data control and ownership decentralization, primarily for privacy and efficiency. This includes healthcare, \cite{agbo2019blockchain, mcghin2019blockchain}, supply chain \cite{durach2021blockchain, wu2019data, sund2020blockchain}, decentralized finance (Defi) \cite{werner2021sok, siyal2019applications}, governance \cite{lumineau2021blockchain}, web browsing, gaming, social media, and file sharing/storage \cite{al2019blockchain}.


Blockchain transactions typically involve digital asset management aligned with business activities. The fundamental transaction type is asset $\transfer$ between accounts, a \textit{native} function in most blockchains. To address the diverse needs of modern applications, blockchains have evolved to include user-designed transactions known as \textit{smart contracts} \cite{szabo1997formalizing}. These contracts execute business operations and adhere to specific conditions. Examples include auction bidding and regulated patient record management. Recent survey \cite{numberOfSC} indicates the existence of over 44 million smart contracts on the $\ethereum$  blockchain alone.

\textbf{Problem:} Smart contracts, despite their flexibility, face adoption barriers due to several issues: (i) They require significant effort in creation and verification, offer limited reusability across platforms, and constrain automatic optimization possibilities. (ii) Vulnerable to user errors and security breaches, they pose financial risks, exemplified by the DAO attack \cite{mehar2019understanding} that resulted in a loss of approximately 3.6M ETH (about \$6.8B). (iii) Many transactional behaviors in smart contracts, embedded in programming structures, remain hidden on the blockchain, hindering their utility in complex data analysis. (iv) Their execution involves higher latency and costs compared to native transactions. The lack of validation semantics for these user-programmed transactions complicates concurrency conflict management, leading most platforms, including Ethereum, to adopt sequential execution, which lowers throughput. 

Declarative smart contracts \cite{chen2022declarative}, domain-specific languages \cite{wohrer2020domain}, and smart contract templates \cite{hu2020smart}  aim to ease creation and verification processes. However, they fall short in addressing performance, throughput, queryability, and other transactional model challenges in smart contracts. 

\subsection{Contributions:} 

 This paper investigates the feasibility and impact of lifting transactional behaviors typically found in smart contracts into the core blockchain layer as \textit{native} transactions. 
 Specifically, we propose:

\begin{enumerate}[itemsep = 1mm, topsep=0pt, partopsep=2pt]
\item a \textit{declarative and typed} blockchain transaction model that includes the novel concept of  \textit{nested blockchain transactions}, as a foundation for modeling transactional behavior on blockchains. \item concrete declarative blockchain transaction modeling of a sample transactional behavior represented in many smart contracts of the most popular blockchain application category - marketplaces.   
\item an implementation framework for declarative blockchain transactions that builds on $\bigchaindb$ blockchain database's architecture \cite{bigchaindbweb}, extending its transaction modeling and validation infrastructure.  
\item a comparative performance and usability evaluation of the declarative transaction model vs. the smart contract model using $\ethereum$ smart contracts as the baseline. The evaluation results demonstrate that the declarative transaction method significantly outperforms smart contracts, achieving improvements by a factor of 635 in latency and a minimum of 60 in throughput.

\end{enumerate}

The rest of the paper is organized as follows: Section \ref{background} provides background information on blockchain native transactions, smart contracts, and $\bigchaindb$. Section \ref{approach} introduces the formal blockchain transaction model and novel concepts of \textit{Non-nested} and \textit{Nested} transactions. Section \ref{implementation} provides implementation details of the concepts presented in Section \ref{approach}. Section \ref{related work} reviews the literature on the topic, while Section \ref{evaluation} reports on the comparative experiments conducted to evaluate our system and smart contract. Finally, we conclude the paper with a summary in Section \ref{conclusion}.

    
     
    
    


\section{Motivation and Background}
\label{background}
\subsection{Smart Contracts in Blockchain Marketplaces}

Most blockchain platforms typically support only basic transactions like $\transfer$, with Ethereum adding more complex types, such as multi-signature transactions that focus more on operational semantics rather than behavior. Consequently, most applications rely on smart contracts to extend functionality, which comes with inherent limitations. For example, in setting up a decentralized marketplace for procurement and supply chain management, smart contracts are needed for actions like posting service requests by buyers or supply bids by providers, involving complex metadata management through user-programmed methods.

\begin{figure}[tb]
  \centering
  \includegraphics[width=\columnwidth]{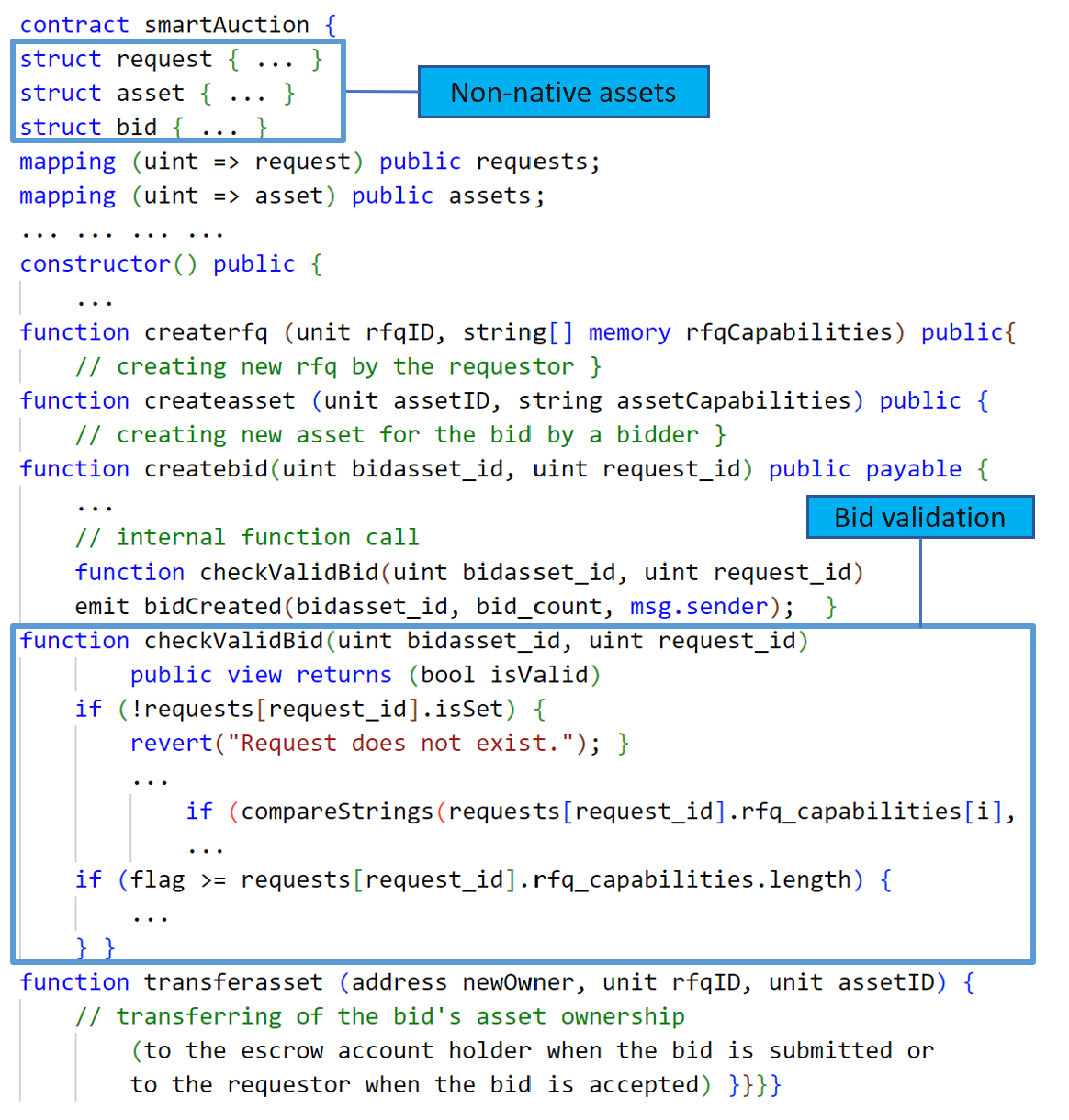} 
  \caption{Smart Contract sample implemented in Solidity}
  \label{fig:sc-auction}
\end{figure}

$\example$. \textit{Buyers can post requests (e.g., for manufacturing services), and providers (e.g., 3-D printer manufacturers) can respond with bids. These transactions involve detailed metadata such as quantity, product type, and deadlines, managed through the $\mathtt{createrfq}$ method for requests and $\mathtt{createbid}$ for bids, which also includes the asset's production capabilities like certifications and work history. This setup mimics traditional auctions where the asset that forms the basis of a bid is some form of payment. Fig. \ref{fig:sc-auction} shows the skeleton of an $\ethereum$ smart contract modeling such a procurement reverse auction marketplace. }

\textbf{Observations:} 

\begin{figure}[tb]
    \centering
    \includegraphics[scale=0.17]{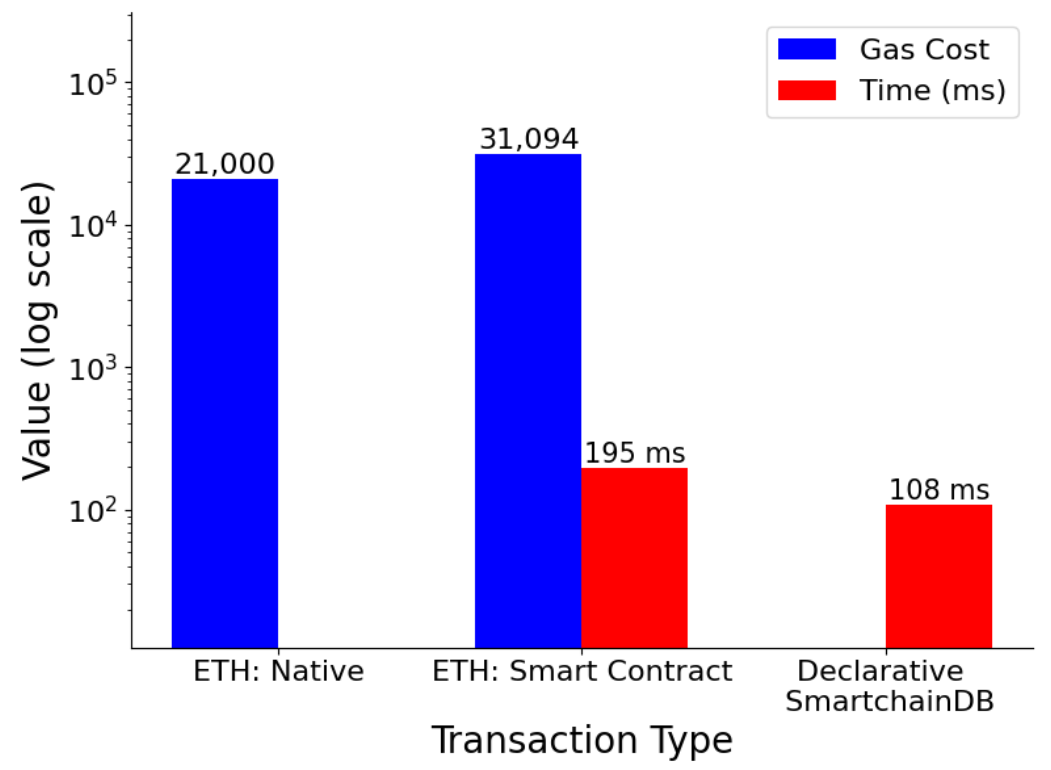}
    \caption{TRANSFER Transaction Runtime and Cost Comparison (Log Scale)}
    \label{transfer-native-non-native}
\end{figure}

Native transactions such as $\transfer$ automatically handle validation against errors like double-spending. However, with smart contracts, developers must manually code such validations, as seen with methods like $\mathtt{checkValidBid()}$. In an auction context, this includes ensuring all non-winning bids are refunded (if escrow deposits were required), verifying ownership of bidding assets, and managing bid withdrawals and deletions by authorized parties only.

Smart contracts also manage a broad range of transaction and asset metadata, which are not directly visible on the blockchain. This includes everything from user content (e.g., documents, audio/videos) \cite{zhu2018digital} to digital twins of physical assets like diamonds \cite{Everledger}, cars and houses \cite{zakhary2019towards} etc.,  even ownership certificates for various physical assets \cite{min2022blockchain, rogerson2020blockchain, park2021effect}. These assets are stored in complex structures that require deep technical knowledge to navigate. In our example, metadata for requests, bids, and their underlying assets are represented as the struct variables, $\mathtt{request}$, $\mathtt{bid}$, and $\mathtt{asset}$ respectively. The mappings of accounts to bids and requests are implemented using program map data structures (e.g. \textit{requests}) and not wallet accounts. Consequently, a query like finding open service requests for 3-D printing manufacturing capabilities may be of interest to 3-D printing manufacturing providers. However, this query involves specifying conditions on the metadata of the service request that are not queryable on the blockchain.
Even more complex queries are critical for supporting tasks like fraud analysis or other business decision-making tasks, but unfortunately, they cannot be supported easily.
Thirdly, the smart contract execution model has more overhead than that of native transactions. An experiment comparing the native $\transfer$ transaction to its smart contract equivalent in Figure \ref{transfer-native-non-native} showed that using smart contracts instead of native transaction primitives increased GAS costs by 40\% in $\ethereum$, reflecting higher transaction latencies and variable execution fees that depends on the contract's runtime behavior. Unlike $\ethereum$'s native transactions, smart contract performance can be unpredictable because it's tied to network conditions rather than fixed processing rules.

\subsection{Rationale for Approach}
Introducing more native transaction types is one way to minimize this dependence on smart contracts. However, several crucial questions must be addressed to accomplish this goal:
\begin{enumerate}[topsep=0pt, partopsep=0pt, itemsep=2pt, parsep=0pt]
\item What new transaction primitives can be added to reduce the burden of always requiring smart contracts in the development of blockchain applications? 
\item How can these new primitives be effectively integrated into blockchains?
\end{enumerate}
To answer (1), we can leverage the fact that the inherent structure of most blockchain applications are marketplaces that facilitate asset trades and tracking. Consequently, common marketplace transactions e.g. buy, sell, bid, etc, are likely to be good candidates for frequently desired blockchain transaction behavior. Indeed, an analytical study \cite{wu2023know}of the smart contracts on $\ethereum$ revealed such smart contract method calls to be dominant. 

 In addressing the integration of new transaction primitives into blockchains, our analysis identifies two main strategies: the imperative and declarative models. Our introductory discussion already highlighted the limitation of an imperative specification model such as smart contracts where the user is responsible for low-level, detailed implementation of transaction behavior. Declarative modeling, similar to that used in relational databases, allows users to define outcomes through constraints rather than detailed processes. This model supports automatic optimization by enabling an optimizer to choose the best execution strategy based on runtime conditions and costs. Also, it is extensible, allowing the combination of simple conditional expressions to form complex ones.
 
For the implementation strategy, blockchain architectures are primarily categorized as native or hybrid. Native blockchains, like $\ethereum$, are built from scratch, focusing on core blockchain features. On the other hand, \textit{hybrid blockchain database systems} \cite{ge2022hybrid} like  $\bigchaindb$ leverage existing database technologies to add blockchain functionalities, offering features such as decentralization, immutability, and controlled asset ownership, alongside high transaction rates and efficient data querying. $\bigchaindb$ integrates blockchain with database capabilities, running on a network where each node operates three services: $\bigchaindb$ server, $\tendermint$, and $\mongodb$. The $\bigchaindb$ server processes transactions, while $\tendermint$, a Byzantine Fault Tolerant engine, handles consensus without mining, using a Proof-of-Stake mechanism. This setup underlines the feasibility of using database-based platforms for a declarative transaction approach, enhancing blockchain’s scalability and performance. Also, $\bigchaindb$ introduces  \textit{blockchain pipelining} technique to improve scalability by allowing server nodes to vote on new blocks before the current block is finalized. Unlike traditional blockchains, where blocks must be sequentially finalized, this approach lets nodes proceed with voting without waiting for a decision on the previous block.

Naturally, the database-based blockchains are a more natural fit for a declarative model since the underlying database likely supports a declarative model. Consequently, we select an approach that builds our declarative transaction approach on top of database-based blockchains, in particular, $\bigchaindb$. 

\textbf{Comment about objective}: This paper does not aim to suggest some minimal set of blockchain transaction primitives. Rather, it aims to demonstrate how declarative blockchain transaction modeling can be achieved and its potential benefits. To this end, it introduces a set of primitives relevant to marketplace applications. The hope is that this set can be extended over time resulting in a corresponding decrease in the dependence on smart contracts, at least for some categories of blockchain applications.

\section{Approach}
\label{approach}

Our approach is to introduce $\smartchaindb$, which is an extension of $\bigchaindb$, with additional blockchain transaction primitives that can be used to specify complex transactional behavior and workflows rather than the use of imperative specifications, i.e., smart contracts. Specifically, we introduce a \textit{"declarative" blockchain transaction} model on which different kinds of blockchain \textit{transaction types} can be based. We then propose some concrete primitive \textit{transaction types} based on the proposed model as well as their implementation strategies. We use decentralized marketplaces as our discussion context, using some marketplace transaction behavior as examples, because of their popularity as blockchain applications. This choice does not limit the generalizability of our approach. In fact, the declarative transaction model we propose is designed to be flexible and adaptable, capable of supporting a wide range of transaction types beyond those presented.


\subsection{Formal Conceptual Model for Blockchain Transactions}
\label{txnmodel}

Our formal transaction model defines key components necessary for a transaction: the asset involved, the participating accounts identified by public keys, the type of transaction, and protocols for automated validation of semantics, such as preventing \textit{double spend} errors in $\transfer$ transactions.
The model is based on the number of sets:

\begin{itemize} [itemsep = 1mm]
    \item  a set $\mathcal{PBPK}$ = \{$\mathtt{pbpk_i}$ = $ < \mathtt{pb_i}$, $\mathtt{pk_i} >$\} of public-private key pairs. The pair $<\mathtt{pb_i}$, $\mathtt{pk_i}>$ represents account/owner $i$. We denote a subset $\mathcal{PBPK}$-$\mathscr{Res}$ $\subseteq$ $\mathcal{PBPK}$ as reserved accounts i.e. system or admin accounts.
    
    \item sets $\mathcal{L}$ – a set of literals. $\mathcal{RK}$, $\mathcal{RV}$ $\subseteq$ $\mathcal{L}$ is a set of string literals that are reserved keywords and values, respectively. We assume a specific subset of reserved values $\mathscr{OP}$ $\subseteq$ $\mathcal{RV}$  that are the names of transaction operations, e.g., $\create$, $\transfer$, and so on. 

    \item a set $\mathcal{S}$ $\subseteq$ $\mathcal{L}$ is a set of strings that are called digital signatures, which are associated with two functions such that given a message string $\mathtt{m: sign(pk, m)}$ returns a signature string $\mathtt{s \in \mathcal{S}}$ and $\mathtt{verify(s, pb, m)}$ is a boolean function that returns $\mathtt{True}$ if the corresponding public key can be used to decrypt the signature and recreate the signed message $\mathtt{m}$. We can also have a more complex string made up as a function of multiple signatures. This is used in the case where an asset is controlled by a group of entities who must sign transactions on the asset. We use $\mathtt{ms_{i,j,k}}$ to denote such a multi-signature string from using signatures generated with private keys $\mathtt{pk_i}$, $\mathtt{pk_j}$, $\mathtt{pk_k}$. 

    \item $\mathcal{AS}$ – the set of all blockchain assets where each blockchain asset $\mathtt{A}$ is a tuple $ \mathtt{< (k_i, v_i), amt >}$ where $\mathtt{(k_i, v_i)}$ is a nested set of key-value pairs such that each $\mathtt{k_i \in }$ \{ $\mathcal{L}$ - $\mathcal{RK}$ \} and $\mathtt{v_i} \in $ $\mathcal{L}$ $\cup$ $\mathtt{A}$ and $\mathtt{amt}$ is a non-negative number of shares that an asset holds. 
    
    \item $\mathcal{T}$ - set of all blockchain transactions
\end{itemize}

$\definition$ \textbf{1. {\textbf{(T\textsc{ransactions})}}}. A transaction $\mathtt{T} \in \mathcal{T}$ is a complex object $\mathtt{<ID, OP, A, O, I, Ch, R>}$ s.t.: 

\begin{itemize} [itemsep = 1mm]
    \item $\mathtt{ID}$ - a globally unique string identifier
    \item $\mathtt{OP} \in \mathscr{OP} $ i.e. the name of transaction operation  
    \item $\mathtt{\{A\} } \subseteq \mathcal{AS}$ - set of assets
    \item $\mathtt{\{O\}}$ - a set of transaction output objects $\mathtt{ \{o_1, o_2, ... o_m \}}$. $\mathtt{T.o_k}$ is used to denote the $\mathtt{k^{th}}$ output of transaction some $\mathtt{T}$ . Since assets can be divisible, the different outputs can hold different numbers of shares of some asset $\mathtt{A_i}$. Consequently, each of $\mathtt{T}$’s output object $\mathtt{o_j}$ is a tuple $< \mathtt{pb_i}$, $\mathtt{A_i.amt}$, $\mathtt{pb^{prev}_i}>$, where $\mathtt{A_i.amt}$ is the number of shares of $\mathsf{A_i}$ associated with the $\mathtt{j^{th}}$ output of $\mathtt{T}$, i.e., $\mathtt{T.o_j[1]}$ that denotes $\mathtt{pb_i}$ is a set of public keys of the owners or controllers of those shares, and $\mathtt{T.o_j[3]}$ that denotes $\mathtt{pb^{prev}_i}$ is a set of public keys of previous owners. 

    \item $\mathtt{I}$ - a set of transaction input objects $\mathtt{\{i_1, i_2, ... i_n\}}$. We use $\mathtt{T.i_k}$ to denote the $\mathtt{k^{th}}$ input of some transaction $\mathtt{T}$. Each input object $\mathtt{i_k}$ is a tuple $<\mathtt{T'.o_b, ms_{u,v,w}}>$, where $\mathtt{T'.o_b}$ is the output that is being "spent" by this input (in this case, the $\mathtt{o_b}$ is an output of some $\mathtt{T'}$) can be referenced by the notation $\mathtt{T.i_k[1]}$ meaning it is the first element of the input $\mathtt{T.i_k}$. $\mathtt{ms_{u,v,w}}$ is the signature string formed from the private keys that should be the signatures of the assets' owners.

    \item $\mathtt{Ch}$ - A set of children transactions. A child transaction is a transaction that depends on the outcome of a preceding parent transaction. It is triggered by the results or changes initiated by the parent transaction, ensuring that subsequent steps align with established rules and maintain workflow integrity.
    \item $\mathtt{R}$ - a reference vector of referenced transactions by their ID. Referencing a transaction differs from spending it, as referencing does not result in the consumption of its output. 
\end{itemize}

$\definition$ \textbf{2.} {\textbf{(N\textsc{ested transactions})}}. Blockchain transaction $\mathtt{T}$ is \textit{Nested} transaction if the following conditions are satisfied:

\begin{itemize}
\item It contains at least one child transaction, denoted as $|Ch| \geq 1$.
\item The parent transaction is considered committed if and only if all its child transactions have been committed.
\item For any parent transaction $\mathtt{T_{parent}}$, there exists at least one child transaction $\mathtt{T}$ within its children set $\mathtt{Ch}$ such that every output of $\mathtt{T_{parent}}$ is included within the outputs of $\mathtt{T}$, expressed as $\forall \mathtt{T_{parent}}, \exists \mathtt{T} \in \mathtt{Ch}: \mathtt{T_{parent}.o} \subseteq \mathtt{T.o}$.
\end{itemize}
Nested blockchain transactions, as defined, incorporate the principle of \textit{eventual commit} semantics, a commitment that is realized through the strategic use of \textit{escrow} mechanisms. This guarantees that a parent transaction is committed only after the successful commitment of all its child transactions.

\subsection{$\smartchaindb$ Transaction Types and Transaction Workflow}
\label{sec-txn-model}

We introduce a novel typing scheme over the set of all blockchain transactions $\mathcal{T}$ that defines a blockchain transaction type $\tau_\alpha$ = $<\mathtt{T}_\alpha$, $\mathsf{C}_\alpha>$ where $\tau_\alpha$ is the subset of transactions in $\mathcal{T}$ that have $\mathsf{OP}$ = $\alpha$ and a set of conditions $\mathsf{C}_\alpha$ defined in terms of a transaction's inputs and outputs. In $\smartchaindb$ there are \textit{Non-nested}: $\create, \transfer, \allowbreak \request, \bidcap$ RETURN, and \textit{Nested}: $\acceptbid$ transaction types.  We say a transaction $\mathsf{T}$ is \textit{valid} with respect to a transaction type $\tau_\alpha$ = $<\mathtt{T}_\alpha$, $\mathsf{C}_\alpha>$ if it meets all the conditions in $\mathsf{C}_\alpha$. For brevity, we present one representative transaction type from the \textit{Non-nested} and \textit{Nested} transaction categories, $\bidcap$ and $\acceptbid$, respectively. The formal models for the remaining transaction types are available in the extended version of our paper \cite{scdb_github}.

$\definition$ \textbf{3.} ($\bidcap$ {\textbf{T\textsc{ransaction type})}}. $\bidcap$ transaction is usually an offer transaction for something being sold or in the context of our procurement example, a $\request$ being made. We make the assumption that typically some asset is used to guarantee a bid and is typically held in some form of escrow account. Given this perspective, a $\bidcap$ can be represented as $\mathsf{\tau_{\bidcap}} = $ $<\mathtt{T}_\bidcap$, $\mathsf{C}_\bidcap>$ where: $\mathtt{T}_\bidcap$ = $\mathtt{<ID, \bidcap, A, O, I, Ch, R>}$.

$\mathtt{ID}$ $=$ $ 95879...,$ $\mathtt{OP}=$ $ \bidcap,$ $\mathtt{A}= \{asset\_id:65be4.. \} $ 

$\mathtt{I}$ $= $$  \{ <KmSd2...,1>, \mathtt{ms_{YM2sd4hn...}}$ \},

$\mathtt{O}  = $ \{ $\:<[7EAsH..], [1]>$\}, $\mathtt{Ch} = \{ \emptyset \} $, $\mathtt{R} = [6ae47... ] $

\vspace{3mm}

\begin{figure}[h!]
  \centering
  \includegraphics[scale = 0.45]{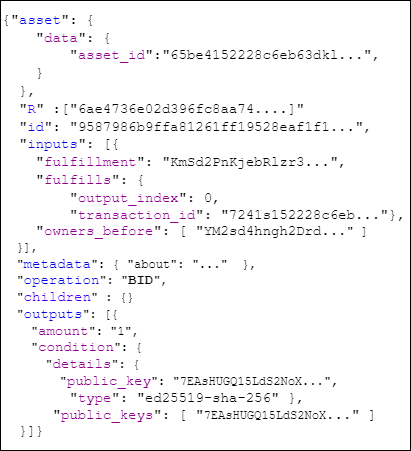} 
    \caption{BID transaction type} 
    \label{bid-tx} 
\end{figure}
For example, the tuple above represents a  $\bidcap$  transaction, with its details illustrated in Fig \ref{bid-tx}. In this $\bidcap$ transaction, the key behavior involves transferring an asset to an escrow account. The escrow account is defined by the output field, where the public\_key of the escrow account is specified. Here’s a detailed explanation of the process:

\begin{itemize}
    \item The input asset, identified by the asset\_id $65be4...$, is transferred to an escrow account. This is achieved by transferring ownership of the asset through the output section, where the public\_key of the escrow account is set as the new owner.

    \item The cryptographic fulfillment, indicated by $KmSd2P...$, ensures that the previous owner's conditions are satisfied before the asset can be transferred. This signature proves that the previous owner authorizes the transfer of the asset to the escrow account.

    \item The output condition and amount for this transaction, $[7EAsH...]$ for the public key and 1 for the amount, define how much of the asset is being transferred and to whom (the escrow in this case).
    
\end{itemize}

This BID transaction is tied to a prior $\request$ transaction, identified by $\mathtt{R} = [6ae47...]$, which references the request being made. This is essential for validating the context of the $\bidcap$, ensuring that the assets and references are properly linked to the request.


A $\bidcap$ transaction has the following set of boolean validation conditions $\mathsf{C}_\bidcap$:

\begin{enumerate}

    \item $\mathtt{|I|} \geq$ 1 i.e. must be at least 1 input object
    
    \item $\mathtt{|R|} \geq$ 1 i.e. reference vector must contain at least 1 element
    
    \item $\exists ! \mathtt{T} \in \mathtt{R}: \mathtt{T.OP} == \request$, i.e., there exists exactly 1 $\request$ transaction in reference vector
    
    \item $\exists \mathtt{i}: \mathtt{T_{\bidcap}.i[1].A.amt} > 0$ i.e. there exists at least one input object with none-null asset
    
    \item $\forall \mathtt{i} \in \mathtt{I},$  $\mathsf{verify(s_i, pb_i, m_i)} == True$  

    \item $ \forall \mathtt{j \in T.o}: \mathtt{T.o_j[1]}$ = $\mathcal{PBPK}$-$\mathscr{Res}$ i.e.  The output of every $\bidcap$ transaction has to be sent to $\mathcal{PBPK}$-$\mathscr{Res}$ account
    
    \item $\mathtt{T.R.A} \subseteq  \bigcup_{j=1}^{\mathtt{|I|}} \mathtt{T.i_j[1].A}$, where ${\mathtt{T.R.OP} == \request}$  The amount of the requested asset(s) must be a subset of the union of input bid assets.

    \item $\forall \mathtt{i} \in [1, \mathtt{|I|}],$  $\mathtt{T.i == T.o_j},$  i.e., every transaction input  $\mathtt{i}$ has to spend some transaction's $\mathtt{j^{th}}$ output
    
\end{enumerate}

$\definition$ \textbf{4.} ($\acceptbid$ {\textbf{T\textsc{ransaction type})}}.  \\ $\acceptbid$ transaction is a \textit{Nested} transaction that takes one or more $\bidcap$ as the parameters. Its semantics is to transfer the winning bid to the requester while unaccepted bids are transferred back the original bidders. 

Formally,  $\mathsf{\tau_{\acceptbid}} = $ $<\mathtt{T}_\acceptbid$, $\mathsf{C}_\acceptbid>$ where \newline
 $\mathtt{T}_\bidcap$ = $\mathtt{<ID, \acceptbid, A, O, I, Ch, R>}$ with the following set of boolean validation conditions $\mathsf{C}_\acceptbid$

\begin{enumerate}[topsep=0pt, partopsep=0pt, itemsep=2pt, parsep=0pt]

    \item $\mathtt{|I|} == n$  i.e. where $n$ is the number of $\bidcap$s for 1 $\request$
    
    \item $\mathtt{|R|} = $ 1 i.e. reference vector must contain exactly 1 element

    \item $\exists ! \mathtt{T_{\acceptbid}} \in \mathtt{R}: \mathtt{T.OP} == \request$, i.e., there exists exactly 1 $\request$ transaction in reference vector

    \item $ |\mathtt{CH}| == |\mathtt{I}|$ number of elements in children set is equal to the number of input objects 

    \item $\forall \mathtt{i} \in \mathtt{I},$  $\mathsf{verify(s_i, pb_i, m_i)} == True$  

    \item 
    $\forall \mathtt{T} \in \mathtt{Ch}: \mathtt{T_{\acceptbid}.o} \supset \mathtt{T.o}$,  i.e., The output of parent $\acceptbid$ is a proper superset of every transaction's output in the children set

    \item $\forall \mathtt{k} \in [1, \mathtt{|I|}],$  $\mathtt{T.i_k[1][1] == \mathcal{PBPK}-\mathscr{Res}},$  i.e. each input has to spend an output of some $\transfer$ transaction that has an account owner  $\mathcal{PBPK}$-$\mathscr{Res}$ 

    \item $ \forall\mathtt{j} \in [1, \mathtt{|O|}] \forall \mathtt{k} \in [1, \mathtt{|I|}]$ : $\mathtt{T.o_j[1][1] == \mathtt{T.i_m[1][3]}} $ where $\mathtt{T.o_m[1][3]} \wedge T.o_j.ID \neq T_{\acceptbid}.A.ID \wedge T.i_k.ID \\ \neq T_{\acceptbid}.A.ID$  is $\mathtt{pb^{prev}_i}$ previous owner of $T.i_m$, i.e., every unaccepted output of $\acceptbid$ transaction must be transferred back to the original bidder

    \item $\exists ! \mathtt{T.o}: \mathtt{T.o[1]} == \mathtt{T_{\acceptbid}.R.o[1]}$, i.e., there exists exactly one output transaction that transfers asset to the requester. 

\end{enumerate}

In a similar way to $\bidcap$ the $\acceptbid$ can be represented in the following tuple form $\mathtt{T}_\acceptbid$:

\noindent $\mathtt{ID}$ $ = $ $ b64c6...,$ $\mathtt{OP}$ $ = $ $ \acceptbid,$ $\mathtt{A} = \{win\_bid\_id:95879... \} $

\noindent $\mathtt{I}$ $ = $  \: \{ $<$HmkC1...,1$>$, $\mathtt{ms_{HmkC1...}},$ $<$MfcDL...,1$>$, $\mathtt{ms_{HmkC1...}} \}$

\noindent $\mathtt{O} = $ \: \{$< $[HmkC1..], [1]$>$\}

\noindent $\mathtt{Ch}$ = \{ $<$[HmkC1..], [1]$>$, $<$[fPjsA..], [1]$>$ \} , $\mathtt{R} = [6ae47.. ] $ 

The tuple above can be described in the following way. For brevity, we will omit previously described general fields like ID, $\mathtt{OP}$ (operation), and $\mathtt{O}$ (output) and focus on transaction-specific fields. The asset $\mathtt{A}$ field anchors the transaction to the specific bid with id $95879…$ that has won acceptance, forming a bridge to the original offer. This transaction includes two Inputs, $\mathtt{Ch}$, $<$[HmkC1..], [1]$>$, $<$[fPjsA..], [1]$>$, each representing the outputs from two different $\bidcap$ transactions for the same $\request$, as indicated in the reference vector R.

Sometimes, complex transaction behavior may require composing multiple transactional primitives into a 
workflow which we define as follows:

$\definition$ 5 {\textbf{(B\textsc{lockchain transaction workflow})}}. Transaction workflow is a sequence of transactions $\mathtt{T_1, T_2, ..., T_n}$ where $\mathtt{T_1}$ is \textit{head} that initiates the workflow and $\mathtt{T_n}$ is \textit{tail} of the sequence. The following condition must be true for a transaction in the sequence: 
\begin{itemize}[leftmargin = 1.2 em, topsep=0pt, partopsep=0pt, itemsep=2pt, parsep=0pt]
    \item $\mathtt{T_1.i} = \emptyset$ Input of the transaction initiating workflow is null.
    \item $\forall \{\mathtt{T_{j}.i} - \{T_1\}\} \exists \mathtt{T.o_k}$ where $\mathtt{T.o_k}$ is committed. The input of any transaction in the sequence, except the \textit{head} transaction,  must come from a committed transaction.  
\end{itemize}

Transaction workflow refers to a series of executions of different types of transactions in a specific order.  The exact number of transaction types involved may vary depending on the workflow. An example can be the utilization of a reverse auction workflow within the context of supply chain procurement. Where the only valid workflows can be, $\create$, $\create - \transfer$, $\create - \request - \bidcap - \acceptbid-\transfer$. This is a multistage process, where one side can $\request$ an execution of a particular item/task and the suppliers can show their interest through $\bidcap$ for this $\request$, and, eventually, if the other side accepts the bid the workflow ends. Other scenarios may include a different number of steps, and/or their structure will be different, but the main point is that they all involve the same primitives.

\section{Transaction Model Implementation}
\label{implementation}

\begin{figure}[tb]
    \centering
    \includegraphics[scale=0.32]{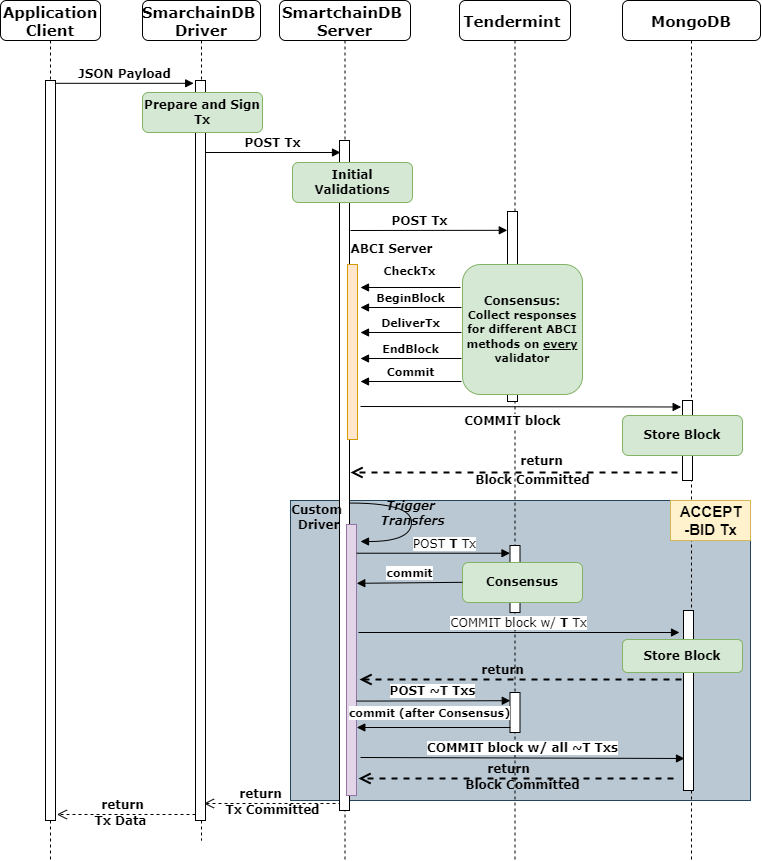}
\caption{$\scdb$ Transaction Life cycle} 
\label{sequence-architecture}
\end{figure}

Our implementation strategy enhances the $\bigchaindb$ platform, modifying its key architectural components, except for the consensus layer ($\tendermint$). Figure \ref{sequence-architecture} depicts the transaction processing workflow and fundamental elements of $\smartchaindb$, our extended system. At the core of our approach, transactions are defined using YAML schemas. Each transaction is validated according to its specific schema type by the $\driver$ before submission to the $\server$. We have enriched the $\server$ with specialized transaction validation algorithms for each type, enabling automatic transaction validation.

In terms of schemas, we have expanded the existing \\ $\bigchaindb$ transaction types $\create$ and $\transfer$, incorporating new components detailed in the transaction model (Section \ref{txnmodel}). This extension also includes schemas for new transaction types like $\request$. On the storage front, the $\mongodb$ collections within $\bigchaindb$ have been adjusted and expanded to support the novel transaction structures introduced in our model. Additionally, the $\server$ component has been fortified with unique \textit{transaction validation algorithms} for each transaction type, facilitating an automated and efficient validation process.

The transaction life cycle begins with the \textit{client} providing a serialized transaction payload in JSON format. Subsequently, $\driver$ utilizes the received payload to generate a transaction by employing pre-existing templates customized to each transaction type and signs it before submitting it to the $\server$ (\textit{"Prepare and Sign"}). 
At this stage, one of the validator nodes is chosen at random to act as the \textit{receiver node}, which is responsible for the semantic validation of the transaction according to the rules for its type. Each transaction has associated  $\mathtt{validateT_{\alpha}}$ method used by the validator nodes at the $\server$ layer, 
 e.g., $\mathtt{validateT_{\bidcap}}$ for the $\bidcap$ transaction. At the network validator node, a transaction undergoes a secondary set of validation checks triggered by the $\mathtt{CheckTx}$ function. This step is implemented to verify that the validator node did not tamper the transaction and to add valid transactions to the local $\mathit{mempool}$. Once a transaction is successfully \textit{committed} on more than 2/3 of validators, the final, third set of validation checks take place at $\mathtt{DeliverTx}$ stage before mutating the state. After accumulating validated transactions, the $\server$ issues a \textit{commit} call to store the newly accepted block to the local \textit{MongoDB}  storage. The $\server$ awaits the response from \textit{MongoDB} about the commit state. Depending on the transaction type, it may end the cycle and inform the \textit{client} about the transaction's status or proceed to the internal \textit{process} from where the response is returned with a successful transaction commit message. The $\driver$ usually attaches a callback to the request, thus, the respective callback method is invoked when the transaction is committed or if any validation error is raised. 
 
  \textbf{Observation}: Blockchain transactions differ from distributed transactions in that they have a transaction "validation" phase done by each peer independently and this phase is delineated from the distributed consensus and commit phases. 
 In the following, we elaborate on the implementation of the transaction validation algorithms for both Nested and Non-Nested transactions. 

\subsection{Implementation of \textit{Non-nested} blockchain transactions}
\label{implementationNonNestedTx}

\begin{figure}[h!]
  \centering
  \includegraphics[scale = 0.28]{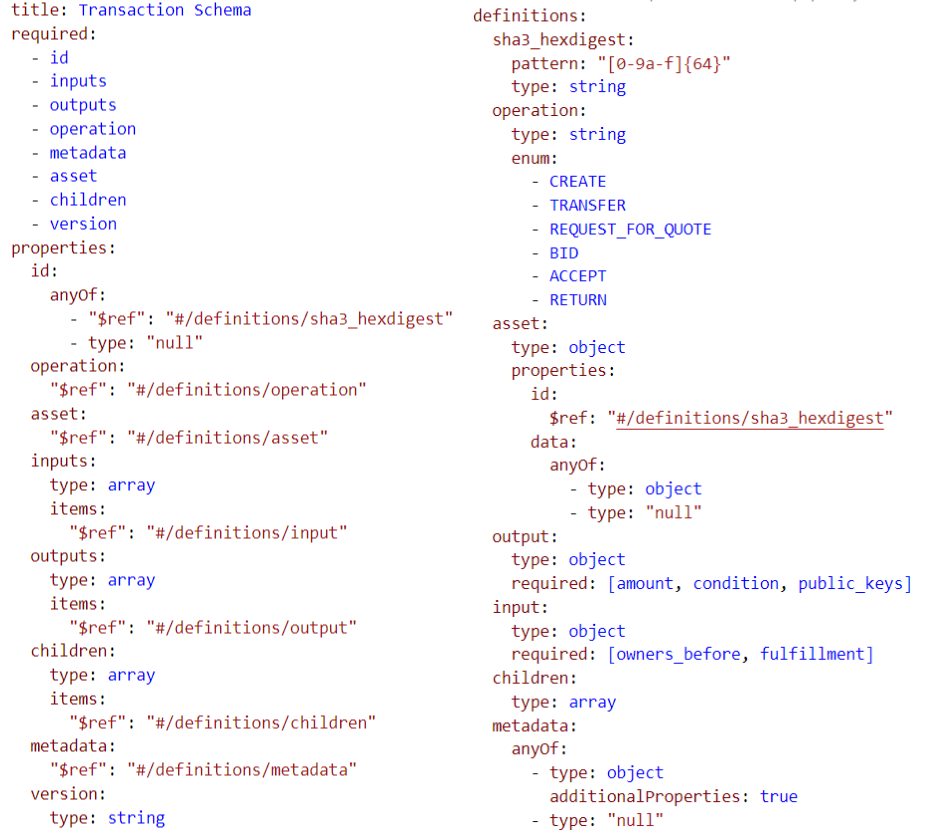} 
    \caption{Transaction Schema in YAML} 
    \label{schema-yaml} 
\end{figure}

Fig. \ref{schema-yaml} presents a segment of a YAML snippet that plays a crucial role in defining the transaction schema in $\smartchaindb$. This schema acts as a blueprint for the formation, validation, and processing of transactions, ensuring conformity to a standardized format. It specifies a rigid structure for transactions, delineating mandatory fields such as id, inputs, outputs, operation, metadata, asset, and version, each with detailed constraints including types, patterns, and additional references within the schema.
\vspace{3mm}

Fig. \ref{schema-yaml} illustrates a portion of the YAML schema that defines the transaction structure in $\smartchaindb$. This schema serves as the foundational framework for creating, validating, and processing transactions, ensuring that each transaction adheres to a uniform format. The schema enforces a strict structure by specifying required fields, including \textbf{id}, \textbf{inputs}, \textbf{outputs}, \textbf{operation}, \textbf{metadata}, \textbf{asset}, \textbf{children}, and \textbf{version}. Each of these fields comes with clearly defined constraints, such as data types, patterns, and references to other components within the schema. This ensures that every transaction meets predefined standards, making it easier to validate and interpret transactions across the system. Additionally, the schema supports flexibility through object references, allowing for modular and scalable transaction definitions while maintaining consistency.

Subsequently, each transaction is subjected to a schema validation method upon arrival at the $\server$. This method employs a schema validation algorithm, described in Algorithm \ref{schema-bid-alg}, that receives a transaction object as input and yields a boolean variable, which signifies the transaction's validity as per the defined schema. The algorithm ensures structural adherence of the JSON transaction payload to the established blueprint.

For example, the $\mathtt{id}$ within the asset definition field imposes constraints that it must adhere to a specific format, as indicated by the reference to a \textit{'sha3\_hexdigest'}, ensuring that each transaction can be uniquely identified and verified. The operation field is restricted to only predefined operations like $\create$, $\transfer$, $\request$, $\bidcap$, etc. This constraint ensures that only allowable transaction types are processed within the $\smartchaindb$ ecosystem. If an operation does not match this predetermined set, it is rejected during schema validation and is prevented from proceeding to the semantic validation phase.


  

  



    
\begin{figure}[tb]
    \centering
    \includegraphics[scale=0.2]{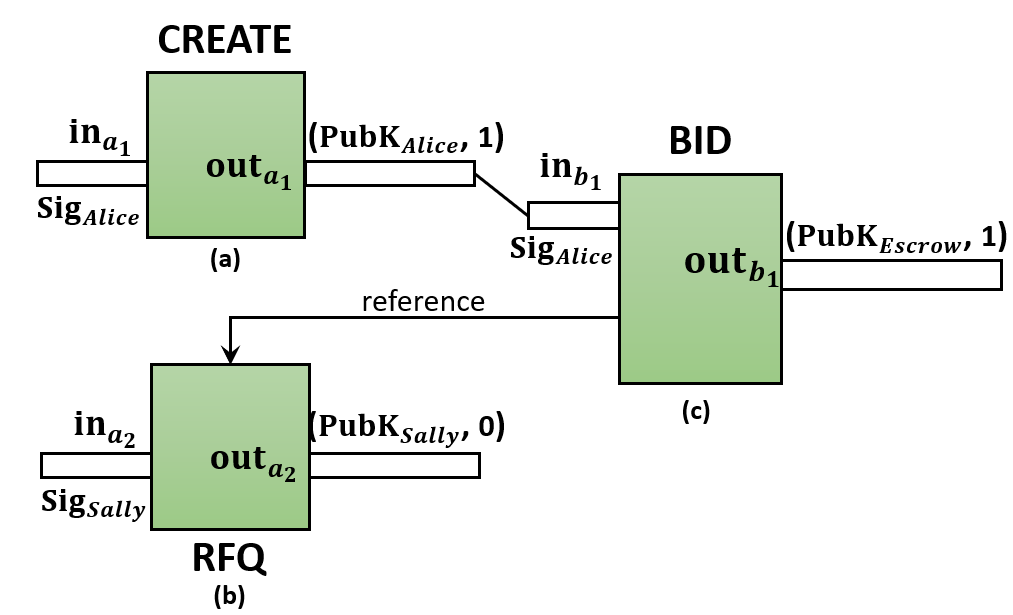}
    \caption{Example of BID } \vspace{2mm}
    \label{create-rfq-bid-example}
\end{figure}

During \textit {semantic validation}, rules about permissions, required dependencies between transactions and conditions about assets are checked. For example, assume Alice responds to a $\request$ for bids by Sally with a $\bidcap$ transaction. Some of the required conditions to check about the bid include (i.) ensuring that Alice owns the asset used to support the bid i.e. she has the permission to spend the output of the $\create$ transaction that created the asset; (ii.) and that the bid is in response to some request and meet some conditions (we ignore additional details).
Fig. \ref{create-rfq-bid-example} illustrates the transaction dependencies (spending and reference) in the example. The permission dependencies for \textit{Alice} are shown by the relationship $\mathtt{PubK_{Alice}}$ and the input signature $\mathtt{Sig_{Alice}}$) while \textit{Sally}'s ownership of the $\request$ transaction is indicated by her signature $\mathtt{Sig_{Sally}}$ on its input. 
The output of $\bidcap$ is owned by \textit{ESCROW} (one of the system accounts) which holds bids until a winning bid is selected. 

\begin{algorithm}[b]
    \caption{$\mathtt{validateT_{\bidcap-schema}}$}
    \label{schema-bid-alg}
\fontsize{9pt}{11pt}\selectfont

  \SetKwProg{Fn}{Function}{ is}{end}
  \newcommand\commfont[1]{\footnotesize\ttfamily\textcolor{blue}{#1}}
  \SetCommentSty{commfont}
  
\KwIn{TxnObject}
\KwOut{Boolean variable}

\SetKwFunction{FbidVal}{$\mathtt{validateSchema}$}
\SetKwProg{Fn}{}{}{}
  \Fn{\FbidVal{$\mathsf{loadSchema(bid.yaml), TxnObject}$}} { }
  
\SetKwFunction{Fbidschem}{$\mathtt{validateTxObj}$}  
\SetKwProg{Fn}{}{}{}
\Fn{\Fbidschem{$\mathsf{asset, TxnObject[asset], data, validateKey}$}} { }  

\SetKwFunction{Fbidschema}{$\mathtt{validateTxObj}$}  
\SetKwProg{Fn}{}{}{}
\Fn{\Fbidschema{$\mathsf{metaData, TxnObject[metaData],}$ $\mathsf{data, validateKey}$}} { }  

\SetKwFunction{Fbidlang}{$\mathtt{validateLanguageKey}$} 
\SetKwProg{Fn}{}{}{}
\Fn{\Fbidlang{$\mathsf{TxnObject, data}$}} { } 

\SetKwFunction{Fbidlang}{$\mathtt{validateLanguageKey}$}
\SetKwProg{Fn}{}{}{}
\Fn{\Fbidlang{$\mathsf{TxnObject, metaData}$}} { } 
 \textbf{return}  $\mathtt{True}$ 
    
\end{algorithm} 
\setlength{\textfloatsep}{0pt}

\begin{algorithm}[tb]
\SetAlgoNoEnd 
  \SetKwProg{Fn}{Function}{ is}{end}
  \newcommand\commfont[1]{\footnotesize\ttfamily\textcolor{blue}{#1}}
  \SetCommentSty{commfont}
\fontsize{8pt}{10pt}\selectfont
  
\KwIn{$\mathsf{rfq\_id, asset\_id, TxnObject, CurrentTxs: List<TxObject>}$}
\KwOut{Boolean variable}

  $\mathsf{RFQTx}$ = \textbf{getTxFromDB}(rfq\_id)\;
  $\mathsf{AssetTx}$ = \textbf{getTxFromDB}(asset\_id)\;

    \If{$\mathsf{RFQTx}$ \textbf{AND} $\mathsf{AssetTx}$ txs are not committed} {
        \textbf{throw} InputDoesNotExistError\;
    }
    \For{every $output$ \textbf{in} $\mathsf{TxnObject.outputs}$}{
        \If{$output.pubKey$ is not $\mathsf{EscrowPubKey}$}{
            \textbf{throw} ValidationError\;
        }
    }
  
    $\mathsf{RequestedCaps}$ = \textbf{getCapsFromRFQ}($\mathsf{RFQTx}$)\;
    $\mathsf{AssetCaps}$ = \textbf{getCapsFromAsset}($\mathsf{AssetTx}$)\;
    \If{$\mathsf{RequestedCaps}$ is not subset of $\mathsf{AssetCaps}$}{
        \textbf{throw} InsufficientCapabilitiesError\;
    }
    
        
    
  \textbf{return} $\FuncSty{validateTransferInputs}$ $\mathsf{(TxnObject, CurrentTxs: List<TxObject>}$)\;

     \caption{$\mathtt{validateT_{\bidcap}}$}
    \label{sem-bid-algo}
    
\end{algorithm}

Algorithm \ref{sem-bid-algo} provides a high-level implementation for $\bidcap$, incorporating semantic validation based on the validation conditions (VC) outlined in subsection \ref{sec-txn-model}. The primary function, $\mathtt{validateBidTx()}$, is executed during the initial validation on the receiver node and twice in the consensus phase on validator nodes (as depicted in Fig. \ref{sequence-architecture}). Initially, a MongoDB query (line 1) retrieves the $\request$ transaction for the specified rfq\_id. The algorithm's first major check (line 4) confirms all transaction inputs, addressing semantics in VC 1-3. Ensuring input transaction correctness, related to VC 4-6, is covered in lines 6-8. A crucial aspect of $\bidcap$ validation, checking if a $\bidcap$ asset meets the required "capabilities", is based on VC 7 and implemented in lines 14-16. Finally, as $\bidcap$ entails aspects of a $\transfer$ transaction, it undergoes additional semantic validation (VC 8) in the algorithm's concluding step (line 13).

\subsection{\textit{Nested} blockchain transactions (NBT)}
\label{implementationNestedTx} 
The traditional “nested transaction” semantics is that a parent transaction is not committed unless child transactions have been committed so that parent transaction blocks on child  transactions. A typical concern is the semantics of nested transactions in the presence of failures. For blockchain contexts, we not only have to worry about being able to recover from failure but also to ensure that security vulnerabilities that allow violation of transaction semantics do not occur due to a failure.

$\example$. Consider a sealed-bid auction with suppliers $\mathtt{Sup_1, Sup_2, ...,}$ $\mathtt{Sup_n}$ submitting bids $\mathtt{T_{B_1}, T_{B_2}, ..., T_{B_n}}$ in response to a $\request$ transaction $\mathtt{T_{\request}}$. The requester initiates an $\acceptbid$ transaction $\mathtt{T_{ACC}(T_{B_1})}$, choosing $\mathtt{T_{B_1}}$ as the winning bid. Correctly, $\mathtt{T_{ACC}}$ should initiate one $\transfer$ of the winning bid to the requester and $\mathtt{n-1}$ $\returnbid$s $\mathtt{T_{R_1}, ..., T_{R_n}}$ back to the original bidders, all from the $\mathcal{PBPK}$-$\mathscr{Res}$ account. These $\transfer$ transactions must be written to the blockchain before committing the parent transaction. Transactions can be executed in \textit{sync} (immediate response before validation) or \textit{async} mode (response after validation confirmation from the $\smartchaindb$ server).

Imagine a scenario where only a subset $s$ of child transactions, say ${\mathtt{T_{R_1}, T_{R_2}, T_{R_3}}}$, completes before a failure occurs. Due to blockchain immutability, transactions in $s$ cannot be undone. A potential issue arises if the $\mathtt{T_{ACC}(T_{B_4})}$ transaction is reinitiated with a different winning bid; it's not a duplicate since it wasn't committed, creating a security risk where the requester might receive both winning bids.

To address this, we propose a \textit{Non-locking} transaction execution approach, allowing the parent transaction to be committed (no lock) to the blockchain even if child transactions are pending. This method enforces 'eventually commit' semantics for the child transactions, ensuring transaction integrity and preventing such vulnerabilities.

\subsubsection{Implementation NLT} Non-locking approach was examined under two scenarios regarding system failures: (1) a positive case without any failures, (2) a case with a possible crash while processing the transaction when more than 1/3 (BFT) of voting power goes offline simultaneously. Under case (1) with no failures, after receiving the transaction payload and performing schema validation, the receiver node logs and sends $\acceptbid$ for the consensus without waiting for the children's transactions to be determined and validated, contrary to \textit{locking} approach. After consensus has been reached, each child transaction, i.e, $\transfer$ is enqueued into a task queue during the commit phase by the receiver node. Multiple parallel workers execute the queued jobs asynchronously. Such an approach enables quick commit of the $\acceptbid$ transaction to the blockchain because it gets committed first, allowing committing all the incoming returns after it in an asynchronous way. Under case (2), when 1/3 of the validator nodes go offline, there are two possible sub-cases: (a) receiver node excluded from the set of the crashed nodes, then the process will resume as soon as sufficient
voting power is attained, and (b) receiver node included to the set of the crashed nodes. The possible node crash times and crash handling techniques under sub-case (2.b) are provided below:

\begin{enumerate}[nosep, leftmargin = 1em]
    \item while processing a parent transaction:
        \begin{itemize} 
        \item{if a crash happens during the initial validation phase, the driver will re-trigger $\acceptbid$ after the timeout interval.}
        \item{if a crash happens on $\tendermint$ in mempool, the election process will be resumed as soon as the quorum of nodes is back online}
        \end{itemize}
    \item  while enqueueing $\returnbid$ transactions: 
    \begin{itemize} 
        \item     enqueue all the $\returnbid$s using the recovery log when the receiver node comes up online
    \end{itemize}

    \item while processing $\returnbid$ transactions:
        \begin{itemize} 
        \item{All the $\returnbid$ transactions already persist in the queue for the execution. $\returnbid$s are sent to a randomly selected validator node to track its commit status and to retry them if needed. Once the chain resumes, they will end up in the mempool and get committed }
        \end{itemize}
\end{enumerate}


\begin{algorithm}[t!]
\fontsize{8pt}{10pt}\selectfont
 \SetAlgoNoEnd 
\KwIn{$\mathsf{rfq\_id, win\_bid\_id, TxnObject, CurrentTxs: List<TxObject>}$}
\KwOut{Boolean variable}
  \SetKwProg{Fn}{Function}{ is}{end}
  \newcommand\commfont[1]{\footnotesize\ttfamily\textcolor{blue}{#1}}
  \SetCommentSty{commfont}

  $\mathsf{RFQTx}$ = \textbf{getTxFromDB}(rfq\_id)\;
  $\mathsf{WinTx}$ = \textbf{getTxFromDB}(win\_bid\_id)\;
  $\mathsf{BidsForCurrentRFQ}$ = \textbf{getLockedBids}(rfq\_id)\;
  
    \If{$\mathsf{RFQTx}$ \textbf{AND} $\mathsf{WinTx}$ txs are not committed} {
        \textbf{throw} ValidationError\;
    }
    \If{signer(Accept-bid) != signer(RFQ)}{
        \textbf{throw} ValidationError\;
    }
    $\mathsf{DuplicateAcceptTx}$ = \textbf{getAcceptTxForRFQ}(rfq\_id)\;
    \If{$\mathsf{DuplicateAcceptTx}$ is in the database}{
        \textbf{throw} DuplicateTransactionError\;
    }
    \If{$\mathsf{WinTx}$ is not found \newline in $\mathsf{EscrowHeldBidsForCurrentRFQ}$}{
        \textbf{throw} ValidationError\;
    }
    \BlankLine
  \textbf{return} $\FuncSty{validateTransferInputs}$ $\mathsf{(RFQTx, WinTx}$)\;
    
    \BlankLine

    \tcp{ Block commit is the final step in consensus}

\SetKwFunction{FCommit}{Commit}
\SetKwProg{Fn}{}{:}{}
  \Fn{\FCommit{$\mathsf{BlockTxs}$: $List<TxObject>$}} {
        \For{every $tx$ in $\mathsf{BlockTxs}$}{
            \If{$tx$ is of type $\mathbf{ACCEPT\_BID}$}{
                $\mathsf{ReturnTxs}$ = $List<TxObject>$\;
                
                r = \textbf{deterRtrnTxs}($\mathsf{WinTx, getPubKey(RFQTx)}$)
                
                $\mathsf{ReturnTxs}$.\textbf{append}(r);\

                \For{every $returnTx$ in $\mathsf{ReturnTxs}$}{
                    $\mathsf{ReturnQueue}$.\textbf{put}($returnTx$)
                }
                \textbf{logAcceptBidTxUpdForRecovery}($tx$, status : \textit{commit})
            }
        }   
    }  
 \caption{$\mathtt{validateT_{\acceptbid}}$}
    \label{sem-accept-algo}
    
\end{algorithm}

\textit{Algorithm discussion.} The gray shaded area in Fig. \ref{sequence-architecture} shows the extra phases required to validate \textit{Nested} transactions using the Algorithm \ref{sem-accept-algo}, that can be divided into two parts. In the first part, parent transaction $\acceptbid$ gets validated according to the conditions from subsection \ref{sec-txn-model} $\definition - 4$. The conditions and errors that can be thrown by this function are readily comprehensible through the pseudo-code provided. For example, if $\request$ and winning $\bidcap$ transactions are not committed or the signer of the $\acceptbid$ transaction is different from the signer of $\request$ transaction, a validation Error is thrown. In the second part, all the appropriate children transactions are determined and written to the blockchain via the invocation of the $\mathtt{commit()}$ method. The $\mathtt{commit()}$ method is called on the receiver node as the last step of the consensus process to trigger children transactions. The function $\mathtt{deterRtrnTxs()}$ determines unaccepted $\bidcap$s for particular $\request$ given the winning $\bidcap$. Once the list of the \textit{n-1} $\returnbid$ transactions has been identified, they all are enqueued to the $\mathtt{ReturnQueue}$ allowing the system to asynchronously send them without blocking the actual flow. To monitor the status of unaccepted $\bidcap$s and to conduct the recovery process, a new collection named $\mathtt{accept\_tx\_recovery}$ was introduced in the \textit{MongoDB} database model. Furthermore, the employed storage model enables reliable queryability facilitating the ability to answer various inquiries.

\section{Evaluation}
\label{evaluation}

In our evaluation, we analyze the performance and usability of blockchain transaction mechanisms by comparing our proposed declarative approach with traditional smart contracts. The usability of smart contracts often requires significant software development expertise, thus posing challenges for contract owners and users. Conversely, our declarative method simplifies the transaction specification process, which is crucial in environments where ease of use is prioritized.

We selected a reverse auction marketplace as our application context due to its involvement of varied transaction types and complexities, such as nested transactions and escrow accounts. These features are common in many practical applications and provide a robust framework for evaluating our system.

Our evaluation focuses on the validation phase of blockchain transactions, which includes the following components:
\begin{itemize}
    \item \textbf{Validation phase}: Each peer that hosts a full copy of the blockchain database validates the transaction.
    \item \textbf{Consensus phase}: A distributed agreement across peers is sought about the transaction's validity.
\end{itemize}

\subsection{Setup}
\label{setup}
\subsubsection{\textit{Experiment environment}}
\label{experiment-environment}
\hfill

\begin{figure*}[t!]
\begin{minipage}[t]{0.9\textwidth}
\centering
\captionsetup[subfigure]{justification=centering}
\subfloat[Latency of $\request$ and $\create$]{\includegraphics[width=0.32\linewidth]{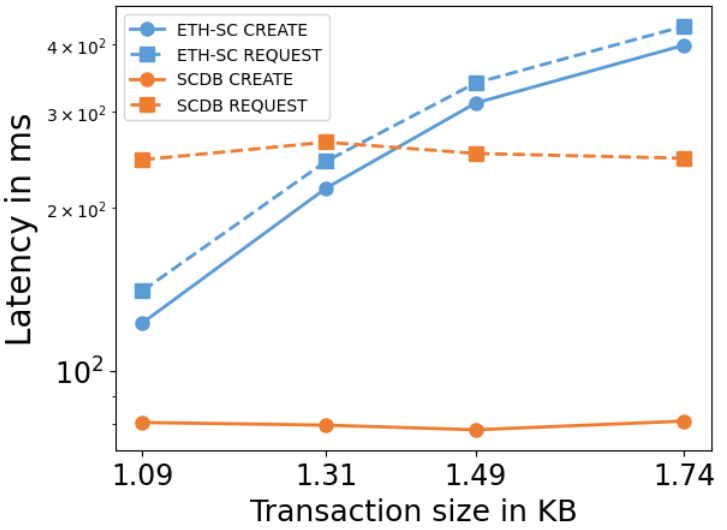}
\label{create_rfq_size}}
\hfill
\subfloat[Latency of $\bidcap$ and $\acceptbid$]{\includegraphics[width=0.32\linewidth]{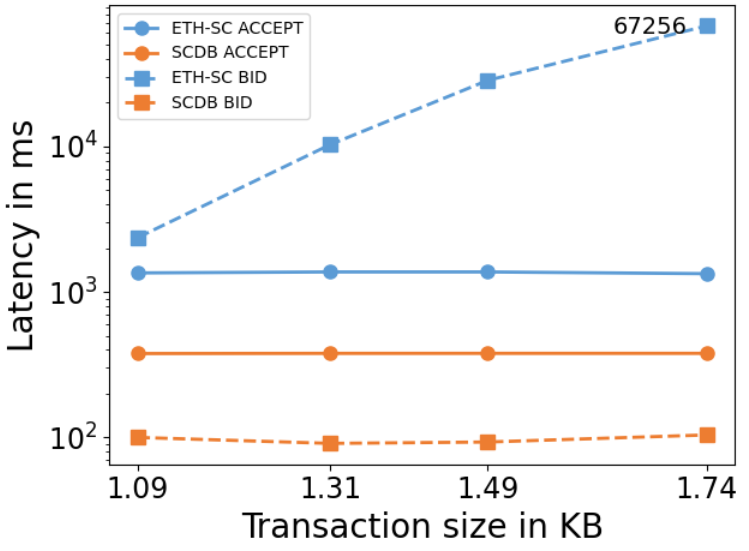}
\label{accept_bid_size}}
\hfill
\subfloat[Throughput]
{\includegraphics[width=0.32\linewidth]{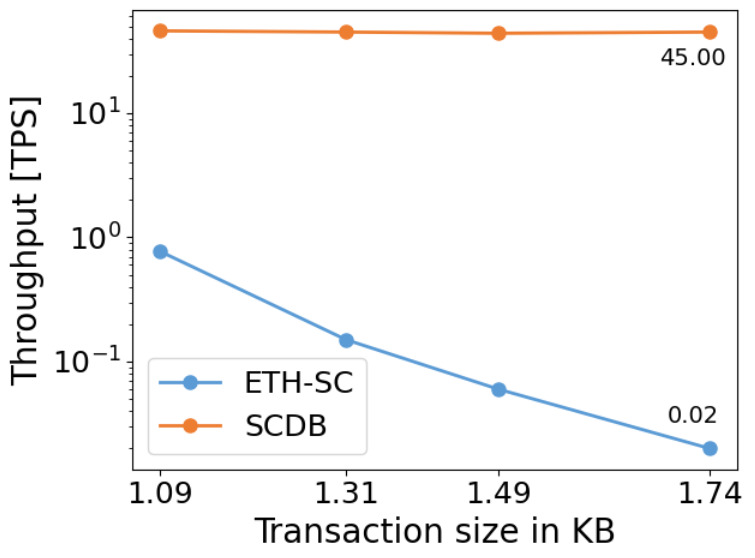}
\label{throughput_size}}
\caption{The Effect of Transaction Size}
\label{effect-of-txs}
\end{minipage}
\end{figure*}

\begin{figure*}[t!]
\begin{minipage}[t]{0.9\textwidth}
\centering
\captionsetup[subfigure]{justification=centering}
\subfloat[Latency of SCDB transaction types]{\includegraphics[width=0.32\linewidth]{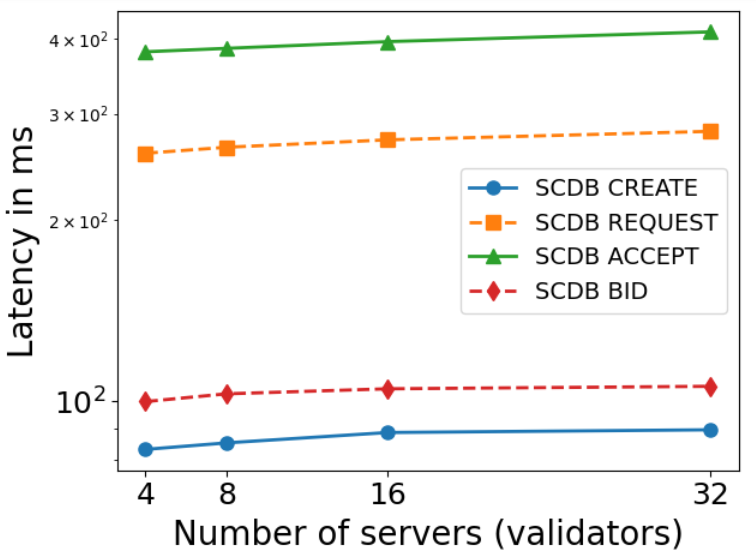}
\label{effect_servers_scdb}}
\hfill
\subfloat[Latency of ETH-SC transaction types]{\includegraphics[width=0.32\linewidth]{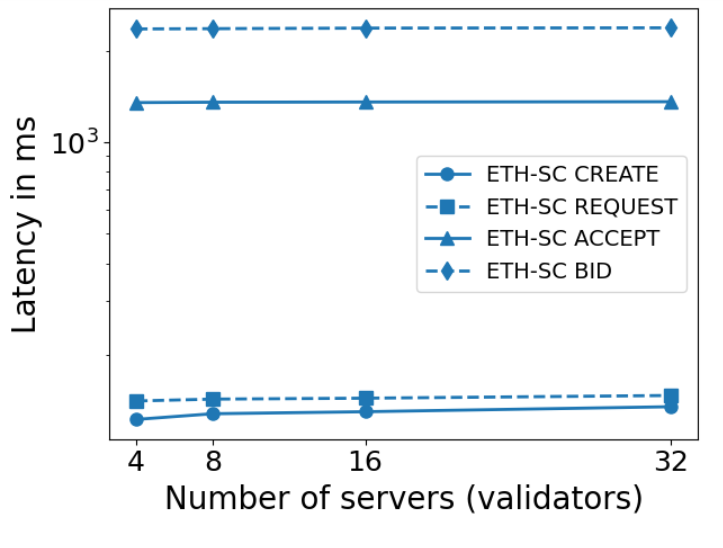}
\label{latency_eth}}
\hfill
\subfloat[Throughput]
{\includegraphics[width=0.33\linewidth]{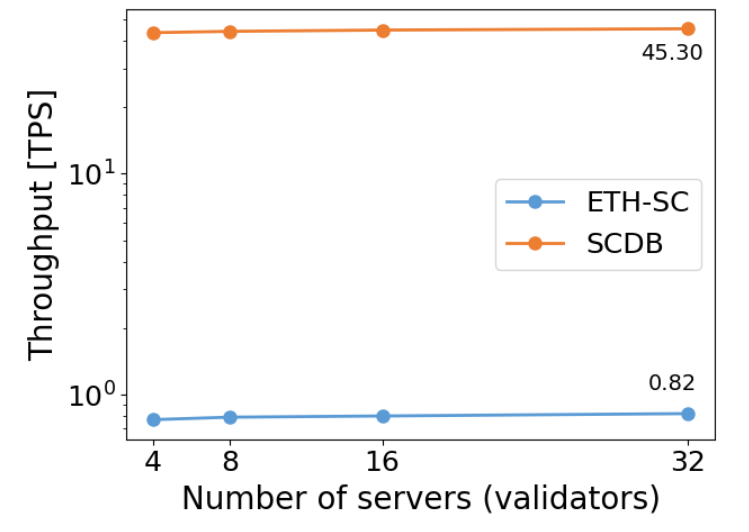}
\label{throughput_cluster}}
\caption{The Effect of Cluster Size}
\label{effect-of-txs}
\end{minipage}
\end{figure*}

The experiments were run on Digital Ocean Cloud using virtual machines (VMs) under Ubuntu 20.04 (LTS) x64 operating system with 8 vCPUs, 16 GB of RAM, and 200 GB of SSD storage. The number of VMs utilized varied depending on the experiment.

\subsubsection{Implementation of Approaches}
\label{smart-contract-implementation}
\hfill


Smart Contract Implementation: For our reverse auction marketplace contract (ETH-SC), we employed Solidity, Ethereum's Turing-complete, statically-typed, and compiled language designed for smart contract development. Our implementation incorporated standard data structures like \textit{struct} to manage user-defined assets, including bids, with transactional functions defined as methods within the contract.

We utilized the Truffle framework \cite{truffle} for automated testing and deployment of the smart contract in JavaScript. To thoroughly test the consensus mechanism in a multi-node environment, we integrated Quorum, an enterprise-focused version of Ethereum. Quorum \cite{goquorum} allows for a permissioned blockchain network that uses customizable consensus mechanisms, making it suitable for our needs.

\textit{Quorum Integration for Consensus Testing:} Our Quorum test network included at least 4 nodes, each on separate VMs, to simulate a distributed environment. Using the Istanbul Byzantine Fault Tolerance (IBFT) protocol, the network ensured finality and low latency, requiring agreement from at least $\frac{2n+1}{3}$ of the nodes for consensus.

We deployed the reverse auction smart contract on this network and conducted multiple transaction rounds. Quorum fully enforced the protocol, offering a realistic assessment of transaction throughput and latency. This setup allowed us to measure the consensus overhead and compare the performance of our declarative approach with traditional smart contract execution, highlighting the trade-offs between usability and the operational costs of maintaining consensus.



For our declarative transactions approach, we leveraged \\ $\smartchaindb$-$\server$, implemented in Python, alongside \\ $\smartchaindb$-$\driver$, developed in Java. Notably, Python, being dynamically typed and interpreted, contrasts with Solidity's compiled nature. Our setup involved a network configuration of a different number of $\server$ clusters, incorporating a consensus protocol. This integration introduces various overheads, including computational, bandwidth, and latency considerations, among others.

\subsubsection{\textit{Workload}} \hfill
\label{dataset}


Blockchains don't have a standard transaction size, so comparing Blockchain X's throughput with Blockchain Y's throughput isn't straightforward because transaction sizes can differ significantly. Consequently, transactions of larger size may require a longer duration for disk writing operations.In our study, unlike \textit{Ge et al., 2022} \cite{ge2022hybrid} that used established benchmark YCSB \cite{cooper2010benchmarking} to evaluate the performance of hybrid blockchains, we recognized the critical role of transaction validation semantics in blockchain performance, including aspects like access rights, asset conditions, and transaction dependencies. This complexity extends beyond simple $read$ and $write$ operations, especially in smart contracts, and requires a more complex workload.

 To accurately evaluate smart contracts, we devised a synthetic workload generator tailored for the declarative transaction approach. This generator creates synthetic payloads varying in data size across different transaction fields. We have sent 110,000 transactions to each system comprising of $\create$: 50,000, $\bidcap$: 50,000, $\request$: 5000, $\acceptbid$: 5000.
 


\subsubsection{\textit{Metric calculation}} \hfill
\label{dataset}

Transaction latency was computed by measuring the time elapsed from the moment the transaction was received to its final commitment.

Throughput was calculated by counting the number of transactions that were successfully committed within a time frame, defined as the interval between the reception of the first and the commitment of the last transaction.

\subsection{Experiments and analyses}
\label{experiments-and-analyses}


The experiments simulate a reverse auction workflow within the manufacturing domain. We conducted four sets of experiments:

\begin{itemize}
    \item Experiment 1: Aimed to evaluate latency and throughput by varying transaction sizes in both systems. The cluster of four nodes was used for both systems.

    \item Experiment 2: Involved a various network size of $\server$ validator nodes to simulate real-world scenarios, evaluating how well the system scales, focusing on throughput and latency across the cluster.
\end{itemize}




\subsubsection{Experiment 1 - Latency and Throughput Analysis with Varied Transaction Sizes} To assess the average latency and throughput, we put a list of strings of various sizes in the \textit{metadata} of $\request$ and $\create$ transactions representing digital manufacturing capabilities being requested and created respectively.

 The data, illustrated in Figs. \ref{create_rfq_size} and \ref{accept_bid_size}, reveal that transaction size had minimal impact on the latency in SmartchainDB (SCDB), remaining nearly constant across all transaction types. Conversely, Ethereum-based Smart Contracts (ETH-SC) exhibited an increase in latency for $\create$ and $\request$ transactions as the transaction weight increased, with latency for $\create$ transactions becoming nearly five times, and for $\request$ transactions, twice that of SCDB. Additionally, the latency for $\bidcap$ transactions in ETH-SC showed substantial growth with increasing transaction size; at 1.74 KB, ETH-SC's latency was 635 times higher (66.43 seconds) compared to SCDB's 0.104 seconds. For $\acceptbid$ transactions, latency remained stable in both systems, although ETH-SC was over four times slower than SCDB.

Furthermore, results in Fig. \ref{throughput_size} indicate that SCDB's throughput stayed consistent despite the growing size of transactions. A notable observation was the inverse relationship between asset size and throughput in ETH-SC, where throughput decreased from an initial 0.72 transactions per second (tps) to 0.02 tps by the end of the experiment.

\textit{Analysis} SCDB vs. ETH-SC: SCDB leverages $\bigchaindb$'s execution architecture, which enhances transaction processing through efficient indexing for database queries, built-in caching for quick data access, and pipelined execution. These features mitigate the transaction payload size's impact on latency. Conversely, in ETH-SC, we observed a consistent rise in latency across all transaction types with an increasing number of transaction size, suggesting scalability issues under heavier workloads. This escalation, particularly for $\create$ and $\request$ transactions (Fig. \ref{create_rfq_size}), ties back to the smart contract's storage structure, comprising a vast array of $2^{256}$ slots. For dynamic data structures like \textit{mappings}, Solidity's hash function computes storage locations, but each map item's retrieval takes $O(n)$ time. Additionally, the complexity of smart contract logic exacerbates latency and throughput issues. The quadratic time complexity ($O$($n^2$)) for $\bidcap$ transactions results from a nested loop comparing each $\create$ asset capability with every $\request$ capability to validate $\bidcap$s. The validation also employs a costly $\mathtt{compareStrings()}$ function in terms of GAS usage. 





\subsubsection{Experiment 2 - Analyzing Impact of Cluster Size on Latency and Throughput.} 
This experiment assessed how the number of validator nodes in the cluster affects latency and throughput in both SCDB and ETH-SC. Throughout the experiment, the transaction size was kept constant at 1.09KB to ensure consistent conditions for evaluation. As shown in Figs. \ref{effect_servers_scdb} and \ref{latency_eth} , despite the increased complexity and number of validators, the latency for various transaction types remained relatively stable for both SCDB and ETH-SC across increasing numbers of validator nodes (from 4 to 32). While adding more validator nodes typically introduces more communication overhead in decentralized networks, the results indicate that ETH-SC's latency does not significantly increase as more nodes are added. This could be due to the efficient finality properties of the IBFT consensus mechanism, which ensures low-latency agreement among nodes. However, despite the stable latency across varying node counts, the baseline latency for ETH-SC is still significantly higher compared to SCDB, particularly for BID and REQUEST transactions.

As depicted in Figure \ref{throughput_cluster}, throughput shows a slight steady increase from 43.5 TPS with 4 nodes to 45.3 TPS with 32 nodes. This incremental improvement in throughput can be attributed to the system's ability to leverage blockhain pipelining technique,which enhances scalability during the voting process for new blocks. With more nodes available, SCDB can distribute the workload more effectively, allowing multiple transactions to be processed simultaneously across different validators. While adding more nodes generally improves throughput, it also introduces potential challenges. Typically, increasing the number of nodes leads to more communication and data exchange among validators, which can slow down the consensus process. These factors account for the steady, incremental increase in throughput, illustrating SCDB's ability to balance performance enhancements with the given complexities.

In comparison, ETH-SC exhibits significantly lower throughput, beginning at 0.77 TPS with 4 nodes and showing no substantial improvement as the cluster size increases. This difference highlights the limitations of traditional Ethereum-based smart contracts in handling high transaction volumes. The overhead of Quorum's consensus mechanism, despite being optimized for permissioned environments, still impacts performance compared to SCDB's more streamlined processing.

\textit{Usability.} To measure the usability of these approaches, the number of lines of code required to implement a new marketplace was counted. $\smartchaindb$ didn't require any user-implemented code, whereas the equivalent smart contract required 175 lines of code to establish one marketplace. 

\section{Related Work}
\label{related work}

Different efforts have been made to address some of the limitations of smart contracts as a mechanism for specifying transaction behavior. 

\textbf{Addressing usability and interpretability challenges:}
Standardized function interfaces or \textit{tokens} such as ERC-721 \cite{erc721} and ERC-20 \cite{erc20} prescribe the minimum set of methods (signatures and behaviors) for specific classes of smart contracts, e.g., fungible tokens. \textit{Smart contract templates} \cite{clack2016smart} are similar in spirit to token interfaces but incorporate methods for linking legal contracts written in prose to methods in a contract so that execution parameters are extracted from the legal prose and passed to the smart contract code to drive execution. \textit{Domain-Specific Languages} (DSLs) such as Marlowe \cite{lamela2020marlowe}, SPECS \cite{he2018spesc}, Findel \cite{biryukov2017findel}, Contract Modeling Language (CML) \cite{Whrer2020DomainSL}, ADICO \cite{frantz2016institutions} are programming languages with limited expressiveness that provide high-level abstractions and features optimized for a specific class of problems (typically in a specific domain such as finance or law).
 DSLs allow the possibility of domain experts rather than programmers to implement smart contracts using graphical user interfaces that can be translated to smart contract code via the DSL.
 However, these techniques still require a non-trivial amount of manual code implementation which is vulnerable to the risk of errors and inefficiencies. Further, being imperative specifications, they are less amenable to querying and analysis. Some contributions have been made in the area of \textit {smart contract code analysis} \cite{grishchenko2018ethertrust, wang2019detecting, gao2019smartembed}, but most have focused on the problem of identifying bugs or attack vulnerabilities. 

 \textbf{Addressing performance challenges:} Several solutions have been proposed \cite{sanka2021systematic} to address the throughput and latency limitations of current blockchains, including sophisticated \textit{consensus algorithms} \cite{gupta2021fault, gupta2021rcc} in $\hyper$, \textit{adjusting block size} which is prone to security vulnerabilities due to the increase in the propagation delay \cite{bitcoinunlimited} and \textit{reducing block data} which provides a limited increase in throughput \cite{lerner2017lumino}. \textit{Sharding} divides the network into different subsets (i.e., shards) and distributes workloads among shards to be executed in parallel. This provides processing and storage scalability, although cross-shared communication overhead is often a major challenge. Further, poor shard design may lead to a 1\% attack and other security issues \cite{hellings2021byzantine, zamani2018rapidchain, wang2019detecting}. Some recent work \cite{tao2020sharding} on a distributed and dynamic sharding scheme that reduces communication cost and improves reliability has been proposed. 
However, these efforts do not directly address the sequential execution of smart contracts adopted by most platforms which limits their throughput.
 

With respect to \textit{parallel execution} of smart contracts, the main challenge is dealing with the conflicts and dependencies between smart contracts, given that they have a shared state.  \cite{dickerson2020adding} propose the use of pessimistic transactional memory systems for concurrent execution of \textit{non-conflicting} smart contracts. They suggest achieving parallelism with lower latency by two steps: first, involving a serializable schedule for miners and, second, executing this sequence of transactions deterministically for parallel validating to avoid the synchronization excessive costs. However, this approach implies that validating should be performed significantly more times than mining. On the other hand, \cite{anjana2019efficient} proposed optimistic transactional memory systems which guarantee correctness through opacity rather than serializability. 
\textit{Speculative concurrent execution} of smart contracts proposed in which transactions are executed in parallel, and if a conflict occurs, by tracing write and read sets, one of the transactions is committed, and the other is discarded to rerun later. Speculative strategies usually perform reasonably well when the rate of conflicts is low \cite{saraph2019empirical, dickerson2017adding, anjana2021efficient}. However, these techniques are still in their early phases and often use read-write sets to define conflicts. Furthermore, empirical results \cite{saraph2019empirical} suggest that this notion might be too aggressive, resulting in many unnecessary conflicts detected, suggesting the need for reasoning about conflicts at a slightly higher level of abstraction. 

Alternative strategies such as aggressive caching and 
parallel validation using validation system chaincode have also been used in $\hyper$ \cite{thakkar2018performance, sukhwani2018performance}. 

\section{Limitations}
\label{limitations}

Declarative transactions excel in scenarios with well-defined and standardized operations. However, they might lack the flexibility required for handling complex, dynamic, or unique transactions that do not fit neatly into predefined patterns. Also, for certain applications requiring fine-grained control over individual steps or transactions, the declarative model might not offer the level of granularity needed.

\section{Conclusion and Acknowledgement}
\label{conclusion}

This paper introduces the concept of \textit{declarative blockchain transactions} and outlines a methodology for implementing it by extending an open-source blockchain database. The objective is to introduce an alternative to smart contracts for representing blockchain transaction behavior due to usability and performance limitations of smart contracts.  Experimental results have validated the rationale behind our approach. 

Our future work will be to generalize our modeling framework further to support more complex transaction modeling, including transaction conditions and compositions. Additionally, we plan to explore modeling concrete transaction types from other blockchain application domains.


Our work was partially funded by a National Science Foundation CSR grant.


\balance
\newpage
\bibliographystyle{ACM-Reference-Format}
\bibliography{scdb}

\end{document}